\documentstyle[11pt,newpasp,twoside]{article}
\def\edcomment#1{\iffalse\marginpar{\raggedright\sl#1\/}\else\relax\fi}
\reversemarginpar

\begin{document}
\title{HI and Galactic Structure}
 \author{Felix J. Lockman}
\affil{National Radio Astronomy Observatory,\footnote{{The National 
Radio Astronomy Observatory is operated by Associated Universities, Inc., 
under a cooperative agreement with the National Science Foundation}}
P.O. Box 2, Green Bank WV 24944,
USA, jlockman@nrao.edu}

\begin{abstract}
HI observations in the 21cm line 
have been a principal tool for investigating the 
large-scale structure of the Milky Way.  This review considers what 
was learned in the first decade after the discovery of the 21cm line,
 and how that knowledge 
 has been expanded and refined in subsequent years.  
Topics include spiral structure, the Galactic nucleus, 
the thickness of the HI layer, and 
affairs in the outskirts of the Galaxy.  New advances in
instrumentation and computing, and a broad attack on problems using 
information from all wavelengths, are likely to keep HI 
 studies of the Milky Way interesting for years to come.
\end{abstract}

\section{Introduction}

Knowledge of the large scale structure of the Galaxy 
has been the single most important
achievement of Galactic 21cm HI studies.  
The nature of Galactic rotation and the thickness of
the  disk were discussed in the second paper ever published on 
21cm observations, mere months after the discovery (Muller \& Oort
1951). The collected information is now vast.  
Major reviews by Burton (1988; 1992), and by Dickey \& Lockman (1990), and
the discussion in  Binney \& Merrifield (1998; hereafter BM98) have 
covered aspects of this topic with varying degrees of completeness. 
For this brief review I will narrow the focus by
considering  what was known of the Galaxy after the first 
10 years of 21cm work and how this knowledge has been
modified,  refined, revised or rejected in the subsequent 40 years.  
The very center of the Galaxy is  a topic of its own and 
 will not be discussed.  
Although the general features of the Galactic system are now
reasonably clear, we still have no consistent and satisfactory understanding of
 many fundamental  aspects of the  HI distribution.  Only a fraction
of the information has been extracted from
 existing HI surveys, and there remains before us the tantalizing 
possibility that with a single new insight, a new feature of the
Galaxy might be pulled  from the data. 

\section{What Oort Knew in 1961}

In April 1961, Jan Oort attended a meeting in Princeton 
 on the topic of the distribution and
motion of interstellar matter in Galaxies, and 
contributed several papers to the proceedings (Oort
1961a, 1961b).  The 21cm line had been detected ten years 
earlier (Ewen and Purcell 1951), but
Oort had been thinking about radio spectroscopy for at 
least a decade before that.  During the 1950s,  Oort 
built telescopes in the Netherlands with students and colleagues, 
parallel work went on in
Australia (e.g. Kerr, Hindman \& Carpenter 1957)   and in a burst of
discovery the Galaxy was mapped.  Seven  years after the first 
detection of the 21cm line, the IAU had a new coordinate system.

By the Princeton meeting in 1961, 
the main elements of the picture were in place, derived
from observations with an angular resolution of no better 
than half a degree and usually 
several times  worse. The
papers at the  meeting are remarkably fresh in
many ways, and illustrate the conceptions (and misconceptions) 
of the time.  Here is what Oort knew then: 

{\bf  HI  is strongly concentrated to a thin layer.}  In the inner Galaxy most  HI lies in a layer
having a FWHM $\sim 220$ pc with little change in this thickness between about
$0.3R_0$ and $0.7R_0$.  But outward from the Sun,  at $R \ga R_0$, the layer thickens
monotonically to the Galaxy's edge. 

{\bf The  HI  layer is not only thin, it is flat.}  In the 
inner Galaxy the mean deviation from a
simple plane is less than 30 pc.  However, beyond the solar 
circle the layer bends into a
systematic m=1 warp whose mean distance from the plane
can be several kpc, many times the layer's thickness.  

{\bf It rotates.}  Galactic  HI  is primarily in circular rotation, with 
noncircular components 
$\la 10 $  km s$^{-1}$ or $<5\%$ of the rotational velocity.  The angular 
velocity of
rotation increases toward the Galactic center.  The rotational velocity 
is almost constant inward 
from $R_0$ to $0.3 R_0$, but then rises to a peak at $R \approx 0.05 R_0$. 

{\bf The Galactic Center is in Sagittarius.}  Galactic rotation 
is approximately symmetric about the radio source Sgr A, but the 
nucleus itself contains many  HI  features which have large
peculiar motions indicative of expansion, for example,
 the 3-kpc arm.  There is evidence for outflow
from the nucleus of $1 M_{\sun} \  y^{-1}$. 

{\bf There are some departures from circular rotation outside the nucleus.}  
High latitude  HI  has
systematic infall, often at large velocity.  The rotation curve 
has small-scale structure, and the
northern rotation curve deviates systematically from the southern curve. 

{\bf HI shows the spiral structure of the Galaxy.}  But the 
northern and southern patterns do not match.

{\bf There is an HI halo.} Considerable  HI  intensities are found 
500 pc from the plane.  These
have no apparent connection with the known high-latitude continuum features.

These items will be reconsidered in light of current knowledge, but first
it is useful to summarize what is now known about HI as a vehicle 
for learning about Galactic structure.

\section{ Living with the Limitations:  Essential Facts about Galactic HI}

The activity of measuring Galactic HI spectra has gone on for 
fifty years, and a number of properties of the data 
relevant to their use for discerning Galactic structure are now evident.  
A more complete discussion of
these issues is in Dickey \& Lockman (1990; hereafter DL90); many have
been known for some time (e.g. Kerr 1969).

{\bf  Galactic HI  is seen in all directions}  and at all 
velocities permitted by Galactic rotation
given the size of the Galaxy. 

{\bf  The total $N_{HI}$ in any direction depends 
 primarily on its Galactic latitude.}  We
live in an edge-on system and 
most of the HI is piled up in the plane.  
Most structure visible in HI at the higher latitudes is 
likely the consequence of specific events in the life of the Galaxy.

{\bf  The line brightness temperature 
does not change much with angular resolution.}   
HI  emission has a
very steep spatial power spectrum $P(k) \propto k^{\alpha}$ with 
$\alpha < -2$ (e.g. Green 1993).  HI is thus
highly correlated from point to
point, with most spatial power on the largest spatial scales.  
HI emission spectra in the Galactic plane change remarkably little 
as the angular resolution is increased
from $36\arcmin$ to $4\arcmin$ (see Figure 3 of Bania \& Lockman 1984).

{\bf  The dynamic range in an  HI  emission spectrum is not large.}  
To first order, most of the
structure in brightness temperature across a spectrum, especially at 
low latitude,  results from
kinematic effects and  thermodynamic structure (e.g. self-absorption).  
``Normal'' emission
features are highly blended, and efforts to isolate individual  HI   
clouds in emission profiles are
usually successful only for clouds which are atypical because of their 
location or velocity.   For
these reasons any quantity which depends on slight 
variations in $T_b$ at different directions, or
at different velocities, is likely not to be well constrained.

{\bf  At low latitudes  the HI optical depth is $\sim 1$ at 
many velocities.}  
Derivation of volume densities is thus uncertain, and quantities like 
the  HI  surface density over the Galaxy 
will also be uncertain because of the 
need for an opacity correction (Burton 1992; Dickey 1993).

{\bf  Velocity crowding can be significant.}  
Galactic rotation projected to the LSR often
compresses a considerable distance into a small velocity interval.  
In the inner Galaxy objects separated by 1 kpc along the line of sight 
have radial velocities which typically differ by 
 only 5 km s$^{-1}$.  In the outer Galaxy 
 1 kpc differences may project to $<1$ km s$^{-1}$.  
Random and non-circular motions will 
cause systematic distortions in kinematically derived structures.

{\bf  There are phase transitions in the ISM.}  
In some locations  HI  is being lost through 
ionization or is recombining; in other locations it is 
lost to the formation of $H_2$ or
regained through molecular dissociation.  Tracing the 
Galaxy through observations of HI
alone will surely lead to incomplete and probably misleading 
conclusions.

\section{ 1961 to 2001: Progress, Revisions, and One Dead End}

\subsection{Spiral Structure in  HI}

     In February 1970, a ``Spiral Workshop" was held at the 
University of Maryland to resolve
some of the differences between the pictures that had emerged 
to that date.  The workshop was
attended by Bok, Burton, Kerr, Lin, Lindblad, Mezger, Shu, 
Weaver, Westerhout and other researchers active in the field 
(Simonson 1970; 1971). 

Kerr discussed the  HI arms at length, beginning by saying  
{\it``...21 cm emission is the way
to study the whole Galaxy..."}, then, after presenting  
a map of spiral structure
(Kerr \& Westerhout 1965; Kerr 1969), 
he discussed a series of qualifications which progressively 
undermined whatever sense of 
confidence one might have had in that map.  Acknowledging that 
{\it ``there is so much 21 cm emission that we're confused by it''}, 
and that {\it ``optical investigations tend to disagree with
the 21 cm investigations in many ways''}, Kerr thought that part of 
the trouble might be because {\it ``the apparent rotation curves 
are different on the two sides, in the first and fourth
quadrants''}, though the cure for this was perhaps worse than the 
disease: {\it ``the thing we've
tended to do so far (quite incorrectly, of course) is to use two 
different rotation curves for the two
sides of the Galaxy."}  Even with this ``fudge'' (not the first or 
the last in this business, as we will see),  the Kerr-Westerhout 
map does not look much like a spiral galaxy.  

At the same meeting Burton showed that it was much easier 
to reproduce the observed structure in  HI 
profiles with modest streaming motions (consistent with observations, and
actually predicted by density-wave theory), than with large  
variations in the true  HI  density
(Burton 1971, 1976).  The converse statement though, is what 
doomed HI studies of spiral structure in
the inner Galaxy:  subtle velocity effects can mimic or 
mask large density differences.  To be fair, this problem was known to the 
early researchers, but they must have hoped that the HI 
spiral arms would somehow shine through the uncertainties.  

It is now generally accepted that evidence of spiral structure in 
inner  Galaxy  HI  profiles is confused at best, and  attempts 
to derive the spiral structure of the Galaxy from  HI  spectra have
largely been abandoned.   While HI can contribute to study of 
arms in the outer Galaxy, arms in the inner  Galaxy seem 
better delineated  in  tracers which have a higher
arm-interarm contrast, or are less ubiquitous, 
than HI.   Molecular clouds are one alternative species, 
but these days anyone needing a model of our spiral structure
usually turns to  ionized gas, either  HII regions or diffuse 
electrons from pulsar dispersion measures (Georgelin \& 
Georgelin 1976; Taylor  \&  Cordes 1993). Models of the 
large-scale morphology of the 
Galaxy are tested against the HI data rather than being derived
from the data (e.g. Burton 1976; Englmaier \& Gerhard 1999).

\subsection{The Inner Rotation Curve and the Galactic Center}

A triumph of early HI studies was in establishing the 
location of the Galactic center, whose position can be 
determined to within a degree or so in longitude 
from the kinematic symmetry of profile shapes 
(Oort \& Rougoor 1960; Blitz 1994).
The  HI  over a large area around the Galactic center, however, has
complex kinematics, which for many years was misinterpreted as 
structure in the basic rotation curve. 

At the Princeton meeting 
Oort (1961a) presented a Galactic rotation curve whose amplitude 
fell gently from the Sun
toward the Galactic center, then rose abruptly to a 
narrow peak at  $R \sim 0.5 $ kpc  before 
falling again.  That rotation curve is quite consistent with current models
from $R_0$ down to $R \sim 1$ kpc 
(e.g. Burton 1992; BM98), but the inner peak is now understood as an
artifact of motions induced by a Galactic  bar, 
motions which have about the same
magnitude as circular rotation at 0.5 kpc $\leq R \leq 1$ kpc.   
When the non-circular component is removed, the Galactic rotation 
curve looks similar to that of other systems and is in reasonable 
agreement with the stellar data (Liszt 1992; Burton \& Liszt 1993). 
The rotation curve at $R < R_0$ is thus now fairly well 
determined, at least to within 10 km s$^{-1}$ or so, which is the 
magnitude of motions ascribed to streaming and asymmetries (but see $\S 5.7$).

The ``anomalous'' HI features in the Galactic nucleus discussed in 
Oort's (1961a) review grew steadily in number  
as more observations were made, until by 1977 there were nearly ten
 (Table 2 in Oort 1977).  More were to follow.  There were attempts 
 to explain some of these, especially the 3-kpc arm, in 
terms of stable orbits rather than expulsion of gas 
(e.g. Shane 1972; Peters 1975), but these models fit only part of 
the data and also tended to predict HI features where none were observed. 
Finally, Burton and Liszt used new data and a full 3-d analysis to show
that the kinematics and distribution of {\it all} the HI 
emission (and the molecules) in the inner $\sim 1$ 
kpc of the Galaxy plausibly arose from gas moving in 
closed elliptical orbits in a 
tilted, bar-like system (Burton \& Liszt 1978, 1983; Liszt \& Burton 1978, 
1980, 1996).  The HI features which Oort believed were ejecta from the 
nucleus, as well as much seemingly ``normal'' HI emission in this 
area, were revealed to be projections of the symmetric, tilted, 
elliptical disk.  No net expansion or mass flux was needed.  
A decade later, the bar was observed in starlight 
 at the location and orientation predicted 
from the HI kinematics  (Blitz \& Spergel 1991b; Blitz 1993).

Our current picture of the Galactic nucleus is much more satisfying 
(and far simpler) than the ``patchwork of kinematically and 
morphologically diverse features'' (Liszt \& Burton 1980) 
that Oort thought was required.  Though there remain many puzzles 
(see, e.g., BM98), and the region is quite complex,  progress 
 has been steady and is likely to continue as more elaborate models 
are developed and tested against the data 
(e.g.  Fux 1999; Weiner \& Sellwood 1999).

\subsection{The Surface Density and Size of the Galaxy in  HI}
     The full two-dimensional surface density, 
$\Sigma_H(R,\theta)$, is unlikely to ever be known
well because of non-circular motions and velocity crowding, 
but there is no fundamental reason why the more robust 
$\Sigma_H(R)$ could not be determined with reasonable
accuracy.  The current state of knowledge, however, is  
disappointing though amusing. 
Consider the  $\Sigma_H(R)$ curve given in Binney and
Merrifield's (BM98) excellent book as 
Figure 9.19, attributed to Dame (1993). It appears quite respectable: 
nearly flat over most of the Galaxy at $4\  M_{\sun}$ pc$^{-2}$   
with an exponential decline
 past $R = 17$ kpc. A look back at Dame's paper, though, 
induces some doubt.   His presentation of the  
same  data has the following notes: {\it ``HI at $R < R_0$ 
 [is] derived from midplane number densities given in Burton \& Gordon 
(1978) and constant HI layer thickness of 220 pc
(FWHM) given by Dickey \& Lockman (1990); values have also been 
arbitrarily scaled by a factor of 2 to approximately match the 
value at $R = R_{\sun}$ (see Liszt 1992)."} 
The provenance of this curve does not 
instill confidence in its accuracy.  

The situation takes an odder turn for  the outer Galaxy, for 
 Dame gives not one, but two possibilities for  $\Sigma_H(R>R_0)$,  
both credited to 
 Lockman (1988) and shown here in modified form as Figure 1.  
 These two estimates were derived (using the old value of $R_0 = 10$ kpc) from 
identical HI data sets, but with different assumptions about the 
outer Galaxy rotation curve: the solid line
is for a rotation curve which is flat at $R \geq R_0$, 
while the dashed line adopts the curve of
Kulkarni, Blitz \& Heiles (1982) which rises linearly by a mere 14\% from $R_0$ to $\sim 2
R_0$ then is flat thereafter. Figure 1 was
originally constructed to make the point that in the outer Galaxy small differences in the
assumptions (specifically in $|dV/dr|$) propagate into large 
differences in $\Sigma_H$, but 
the  size of the Galaxy in HI also depends on the shape of the
rotation curve beyond $R_0$, as does the total HI mass, which 
changes by 20\% between the two models. 

    The sensitivity of the
surface density to the rotation curve was used by Knapp, Tremaine \&
Gunn (1978)  to make inferences about the size of the HI disk and the form of
the curve itself.  They were able to limit the size  because 
HI emission does not 
extend to the maximum velocity permitted by Galactic rotation. 
But the parameterization of the rotation curve that they adopted 
would not allow for the result given by the dashed line.

It has been by  argued that the rotation curve is probably flat or 
slightly declining beyond the Sun (Binney \& Dehnen 1998), but  
if $\Sigma_H(R)$ is also to be flat or declining for all $R >
R_0$, then an increase in the circular rotation velocity  
between $R_0$ and $\sim2R_0$  seems
required, exactly as a naive interpretation of the rotation curve
data  would  suggest (Wouterloot et al. 1990; Merrifield 1992). 
While it seems certain that  the vast majority of Galactic HI lies in the
outer Galaxy at $R > R_0$, the actual amount out there is not 
so  well known.

\begin{figure}
\plotfiddle{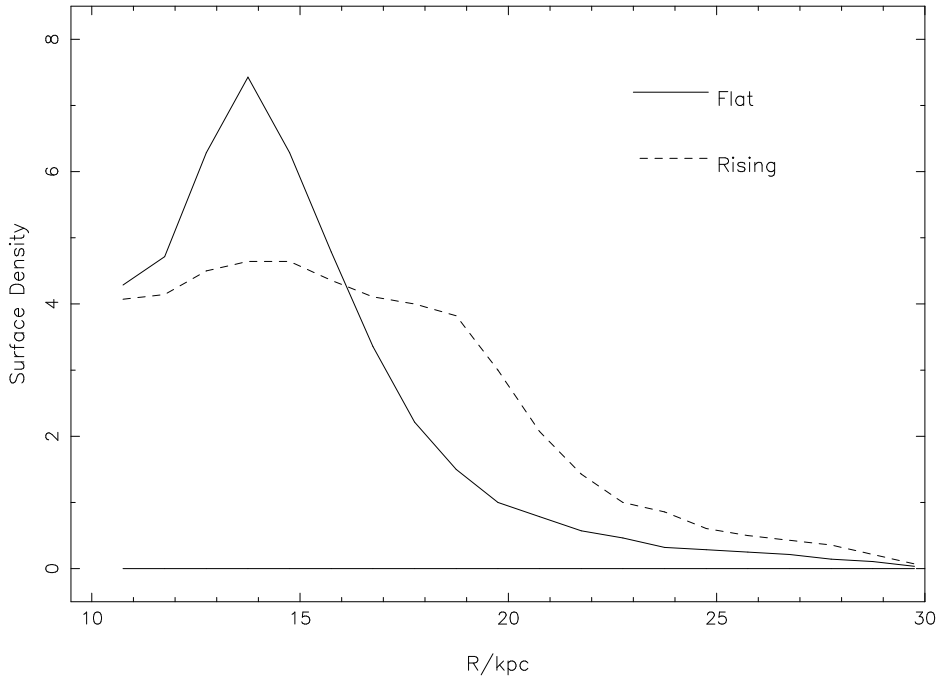}{3.4in}{-90}{50}{50}{-0180}{290}
\caption{The HI surface density in the outer Galaxy is extremely sensitive
to the rotation curve used to derive it.  The curves show the implied 
surface density $\Sigma_H(R)$ in $M_{\sun}$ pc$^{-2}$ 
for data from the Weaver \& Williams (1973) survey 
when analyzed using a flat rotation curve (solid line) where 
 $V_{cir} = V_0$ for $R \geq R_0$, and for a rotation 
curve which rises by about 2 km s$^{-1}$ kpc$^{-1}$ from $R_0$ to $2R_0$ 
then is flat thereafter (dashed line). The two rotation curves differ by
only a few percent in amplitude at $R= 12-15$ kpc.   
The scale is based on the old $R_0=10$ kpc (adapted from Lockman 1988). }
\end{figure}

\subsection{Scale Height and Mean z in the Inner Galaxy.}

The high degree of flatness of the  HI  layer in the inner 
Galaxy was noted in the first surveys, and the HI plane was taken to 
be the best reference for a ``Galactic plane'' (Blaauw et al. 1960). 
But interestingly, although 
 the HI layer at $R<R_0$ is thin and flat,  it is not completely thin, 
and it is not completely flat: there is some residual waviness of 
the mean about the plane (Gum, Kerr, \& Westerhout 1960).  
The deviation of the mean layer from $z=0$ is only a few tens of pc 
(Quiroga 1974), but the  phenomenon is  significant because the 
deviations are to some extent coherent, and are shared by
every Population-I type species in the inner Galaxy,  suggesting 
that the gravitational potential itself may be distorted 
(Lockman 1977; Spicker \& Feitzinger 1986).  There is as yet no
explanation for the ``corrugations", as they are usually called, 
but they have been detected in
other galaxies (Florido et al. 1991).

  The scale height also has interesting structure.  
Oort (1961a) reported that it was approximately
constant over much of the inner Galaxy, though 
somewhat smaller at small R and somewhat larger near the Sun. 
This has been a matter of theoretical concern, for the scale
length of the stellar disk is only a few kpc, and unless supporting forces somehow compensate
exactly for
the change in the local gravity, the HI layer should thicken significantly from $0.25R_0$ to
$R_0$ (see Ferrara 1993 for a discussion of this point, 
and an ingenious solution).  

In the inner Galaxy the tangent points are unambiguous locations for measuring
the thickness of the layer.  Malhotra (1995) modeled 
 the emission at the tangent points in the first and 
fourth quadrants with a single HI component and  derived a scale height 
which is approximately constant between $0.25R_0$ and 
$0.7R_0$,  then rises slightly approaching $R_0$.  Although 
Malhotra ultimately interprets the data somewhat differently, her Figure 6  
indicates that Oort's understanding is still accurate.   

The vertical structure of HI in the inner Galaxy has further interesting
complexities.  Figure 2 shows a velocity-latitude diagram for HI at 
longitude $10\deg$ (all v-b diagrams in this
paper use data from the Leiden-Dwingeloo survey of Hartmann 
\& Burton 1997). The tangent point here has  $R = 0.17R_0$, 
certain enough, but the velocity at which the tangent-point 
gas manifests itself
depends on the rotation curve, the velocity dispersion, 
and the spatial arrangement of the HI.  I have calculated the vertical 
FWHM of the HI emission at three velocities and the results are indicated by
barred vertical lines at 180, 190 and 200 km s$^{-1}$.  At this
tangent point, the FWHM is not very sensitive to the velocity at which 
it is measured, having  a value of 120 pc at all three velocities (and 
note the ``corrugation'' of the HI upward from $b=0\deg$). 

\begin{figure}
\plotfiddle{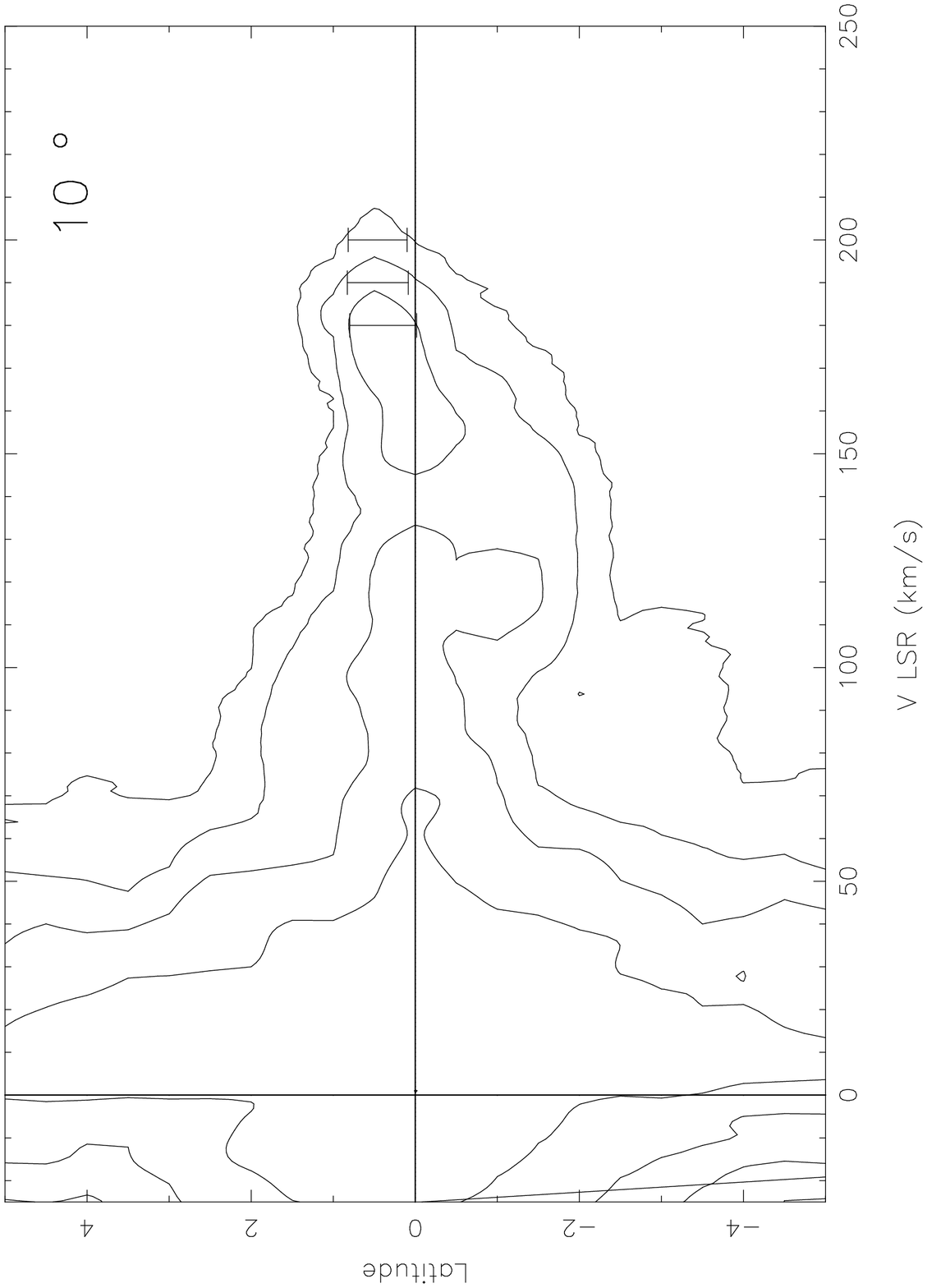}{3.1in}{-90}{45}{45}{-0180}{260}
\caption{Velocity-latitude diagram for HI at longitude $10\deg$.
The vertical barred lines show the FWHM of the emission at velocities 
180, 190, and 200 km s$^{-1}$.  The layer has a thickness of about 
120 pc independent of the exact velocity at which it is measured.
}
\plotfiddle{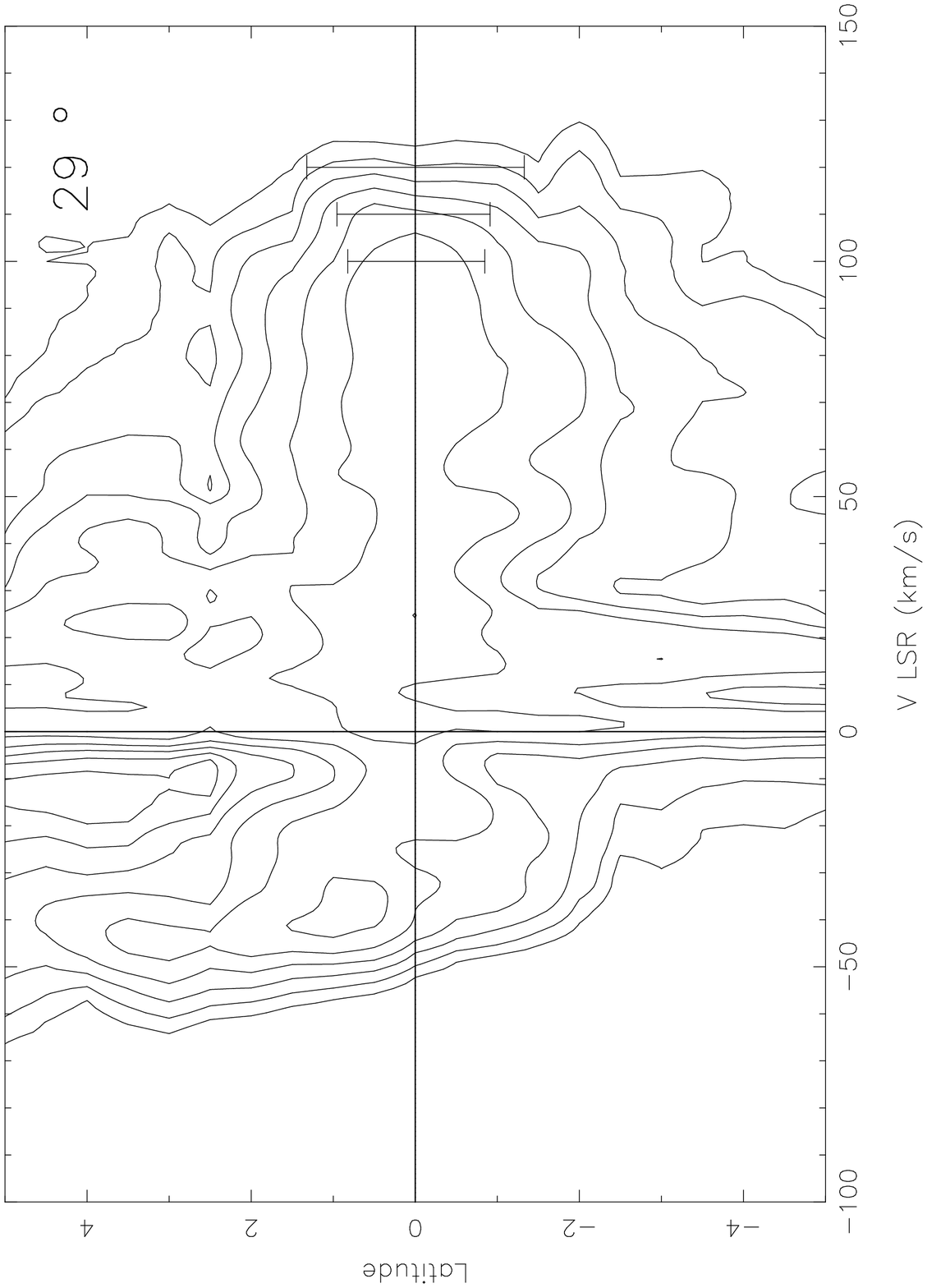}{3.5in}{-90}{45}{45}{-0180}{260}
\caption{Velocity-latitude diagram for HI at longitude $29\deg$.
Vertical barred lines mark the FWHM of the emission at three velocities, and
illustrate how gas in the profile wing
 lies in a much thicker layer than gas at lower velocity.
The FWHM increases steadily from 220 pc to 345 pc between 100 and 
120 km s$^{-1}$, though gas at all these velocities must come from 
nearly the same location. 
}
\end{figure}

Figure 3 shows a similar v-b plot at longitude $29\deg$, 
where the tangent point is at $R = 0.48R_0$.
The FWHM is again calculated at three velocities, but  
here the scale height is very much greater 
in the line wings than in the brighter emission. This 
behavior is seen at many longitudes  and
shows that the Galaxy has an HI component with high
velocity dispersion and high scale height which, at low 
latitudes, stands out only in the line wings 
(Lockman 1984).  The high-dispersion component can be 
traced many hundreds of pc from the
plane and is not encompassed by Malhotra's (1995) single-component
model.  The  gaseous 
``halo'' discussed by Oort (1961b)  has now 
been seen in many species, and in other galaxies, 
and it poses interesting problems  
(Lockman, Hobbs, \& Shull 1986; Savage 1990; Sembach 1997;  
Ferrara 1993;  Sancisi et al.~2001; Benjamin, this volume).

The multiple components  of HI in the 
inner Galaxy are probably responsible for most differences in 
results between various studies 
of the HI scale height, for different methods of analysis  can 
emphasize one component over another. 

\subsection{The Outer Galaxy: Warping and Flaring}

There is considerable current interest in HI in the outer Galaxy, 
stimulated in part by studies of  other
galaxies and  in part by 
the possibility of  using  HI as a tracer of dark matter.  
This brief discussion will not do justice to the current pace of 
research, especially to the extragalactic and theoretical work.

The Galactic warp is the least well-understood major Galactic structure.  
Whereas the flatness of
the HI layer in the inner Galaxy was recognized as having
 ``fundamental physical significance"
(Blaauw et al. 1960) the shape of the warp may be a mere contingency.  
The description of the warp given by Oort (1961a) is still generally correct, 
only the scale has changed with use of newer rotation curves 
(Diplas \& Savage 1991; Voskes \& Burton 1999).  A 
general review is in  Burton (1992).  The
warping begins at about $1.5 R_0$ and is seen in molecular gas 
  as well as  many species in UV absorption lines 
(e.g. Wouterloot et al. 1990; Savage, Sembach \& Lu 1995). 
Emission in the 21cm line remains its best tracer.  
The conclusion that the Sun lies near the line of
nodes of this $m=1$ warp is generally viewed as an unfortunate coincidence and 
not as a misinterpretation of something else
(e.g. Kuijken \& Tremaine 1994). 

The dramatic flaring of the HI layer at $R>R_0$, in contrast  
to the more ambiguous run of scale-height with R in the inner Galaxy,  
greatly simplifies our understanding of it.  As the
surface density of the stellar disk decreases in the outer Galaxy,  the
gas layer should puff up unless its  means of support
also decreases.  This property of HI 
has in fact been  used to measure the disposition of dark
matter in the outer Galaxy by Olling \& Merrifield (2000), although 
in their application they
assumed  total turbulent support for the HI layer, an assumption which has been
roundly criticized (Merrifield 1993) when applied to the solar 
neighborhood (Lockman \& Gehman 1991).

The difficulties in a traditional kinematic analysis, 
and the potential importance of understanding the flaring and warp, 
led Merrifield (1992) to appeal to geometry and symmetry
and derive  the scale height and rotation curve simultaneously from 
the HI data.  This
approach has its own set of biases and may be susceptible to  
any ellipticity in the disk, but offers an independent 
 constraint on Galactic parameters (Kuijken \& Tremaine 1994).

Warps are now known to be extremely common among galaxies 
whose HI extends past the
optical disk (Briggs 1990; Kuijken \& Garcia-Ruiz 2001).  There is a tight 
correlation of HI surface density with
the surface density of dark matter implied from the rotation 
curve (Sancisi 1996;  Hoekstra, van Albada, \&
Sancisi 2001), which has prompted suggestions that the HI is somehow 
coupled to the dark matter, either materially 
(Pfenniger, Combes \& Martinet 1994) or 
epistemologically (e.g. Freeman 2001). 
This area of research is a rare one where Galactic and 
extragalactic studies are congruent.  The region of the warp may be a
 zone of transition between the disk and the halo; between gas in 
near-circular orbits, and high-velocity clouds.

\subsection{ Anomalous Velocity HI -- High-Velocity Clouds}

The sky is falling: the low-velocity HI at the 
Galactic poles is descending gently but systematically  at a 
few km s$^{-1}$, illustrated by Weaver (1974) in 
his  marvelous Figure 9. 
There is also intermediate velocity HI falling 
toward the plane at about $-50$ km s$^{-1}$ 
over a large part of the northern sky (e.g. Kuntz \& Danly 1996).  
Both phenomena may be  local, 
related to the bubble of hot gas around the Sun and the 
relative paucity of HI at high latitude (DL90; Cox \& Reynolds 1987).    
More generally, recent HI surveys show that almost 40\% of  
the sky is covered with anomalous-velocity HI  emission 
($|V_{LSR}| \geq 100$ km s$^{-1}$), whose velocity is unlikely to 
arise solely from  Galactic rotation (Lockman et al. 2002).  
For many years  high-velocity clouds were detectable only in HI. 
They are  a source of interest, confusion and dispute.  
The topic is reviewed by  
Wakker \& van Woerden (1997) and by  Putman in this volume, 
so I will comment only  on the possible connection
between the high-velocity gas and Galactic structure.

\begin{figure}
\plotfiddle{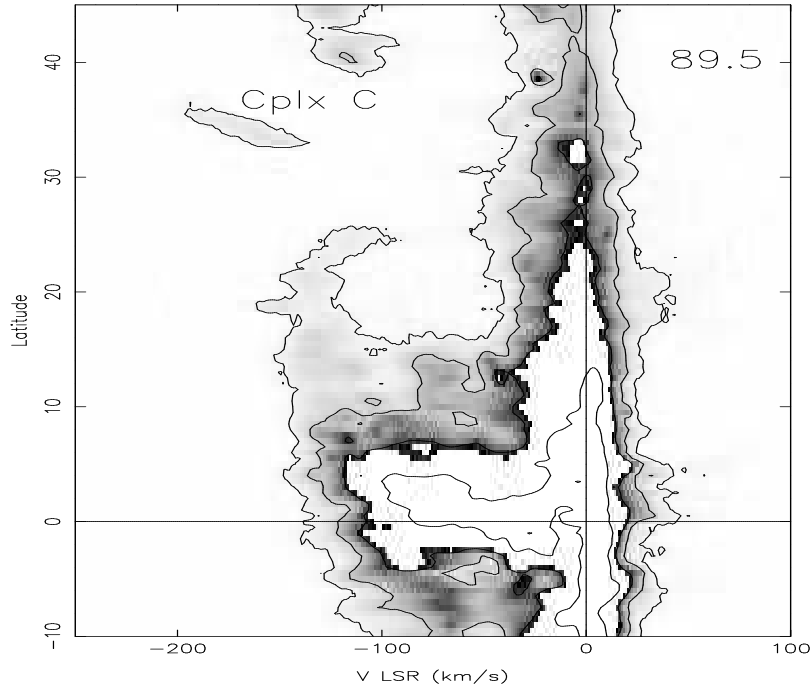}{3.5in}{0}{60}{37}{-175}{-15}
\caption{Velocity-latitude diagram for HI at longitude $89\fdg5$. 
The mean layer warps to higher and higher z with 
increasing distance from the 
Galactic center (to negative velocity) and lies more than $20\deg$ 
away from $b=0\deg$ at its most extreme.  Above the warp, in the same
velocity range, is the group of high-velocity clouds called Complex C.
}
\end{figure}

If the Galactic warp extends far enough from the plane, 
could it be mistaken for a high-velocity cloud?  In the v-b plot of Figure 4, 
distance from the Sun (and from the Galactic center) increases  
 to the left.   The HI contours begin to warp away from $b=0\deg$ 
around $-40$ km s$^{-1}$, and
by $-125 $ km s$^{-1}$ (or $R = 20$ kpc for a flat rotation 
curve and $R_0=8.5$ kpc) the centroid is
nearly five degrees above the plane at an implied $z = 1.5$ kpc.  
The warp then shoots up quite steeply in faint emission at more negative
velocity, and appears to be trying  to connect up with 
 several high-velocity clouds belonging to
Complex C.  The fact that the assuredly ``Galactic" HI emission at 
latitudes $\la 20\deg$  seems to continue toward the high-velocity 
clouds lying at the same
velocity, has led to suggestions that the warp and the high-velocity 
phenomena are the
same, or at least related, though at the sensitivity level of the 
Leiden-Dwingeloo survey there is
no connecting emission.  One subclass of high-velocity clouds 
is generally accepted to be an extension of one of the outer arms 
of the Galaxy (e.g. Habing, 1966; 
Haud 1992), but the link between the
others and the Galactic disk is more obscure and in many 
cases implausible.  

If most of the positive-velocity 
high-velocity complexes at $\ell > 180\deg$ belong to a leading
arm of the Magellanic Stream (Putman \& Gibson 1999; 
Sembach et al. 2001), then the vast majority of the high-velocity HI {\it not} 
associated with the Magellanic Stream has a negative velocity, 
and  Oort's belief that high-velocity clouds are primarily 
infalling with respect to the LSR remains correct.  

Some of the high-velocity complexes have distances which place 
them only a few kpc away and thus in the Galaxy's halo
 (van Woerden et al.~1999).  They must give the Galaxy  a
very ragged edge.  The environment around the Milky Way 
may be not too different from that of 
other interacting systems  where Galactic disks 
swim in a sea of gaseous debris (e.g. Gibson et al. 2001). 

\subsection{ Galactic Asymmetries} 

Gas in a circular orbit around the Galactic center will have a
velocity $V_{LSR} = R_0 sin(\ell)cos(b)[\Omega(R) - \Omega_0]$.
Thus, in the inner Galactic plane, the terminal velocity $V_t(\ell' < 90\deg)$ 
should be of equal magnitude but opposite sign to the 
 $V_t$ at the symmetric position $-\ell'$, because in both directions  it 
 arises from the same galactocentric radius $R = R_0 sin(\ell)$.
It was known in 1961 that the HI  deviates from this 
symmetry at the level of $\pm 10 $ km s$^{-1}$, and that the deviations 
have some large-scale coherence. 
 Simple fixes to restore the symmetry by applying
a radial motion to the LSR (Kerr 1962) do not solve the problem 
completely (e.g. Kerr 1969; Lockman 1988).

If there is an outer edge to the HI disk which occurs at an approximately 
constant $R_{max}$ around the Galaxy, there should be another
 large-scale symmetry 
in the velocity structure of one wing of all HI profiles, 
proportional to $sin(\ell)$.
The true ``outer edge'' of Galactic HI can be fairly irregular without 
distorting the $sin(\ell)$ structure because velocity crowding is so
severe at large $R$ that details a few kpc in size are not
discernible.  But again, the data show clear, systematic deviations 
from the expected behavior (Blitz \& Spergel 1991a). 

Asymmetries in the inner Galaxy are probably a manifestation of the 
bar or other components of  the Galactic potential, 
but the situation at the
Galaxy's edge is less certain.  The  observations may  indicate  
a general asymmetry of the Galactic potential, or give proof that the 
HI disk is not in equilibrium or is extremely lopsided
(Blitz \& Spergel 1991a;  Kuijken \& Tremaine 1994).

\section{ What Next?}

\subsection{More, Better, HI data}

 High angular resolution is not 
always necessary  to  resolve large-scale Galactic features,  which is why most
of what Oort knew in 1961 is still basically correct. 
As Kerr noted {\it ``in more recent years, new surveys with higher
resolving power at Parkes and Green Bank have yielded an enormous 
amount of new information ... but have not yet greatly improved the 
understanding of the basic problems''}  (Kerr 1969).  
But consider: one degree corresponds to
about 150 pc at the Galactic Center, which gives a few, but only a
few, independent points across the HI layer at
most locations in the inner Galaxy, and  resolves the warping, 
 flaring outer Galaxy HI layer only at $90\deg \leq \ell
\leq 270\deg$.   A 100-meter antenna has 
a resolution of $\sim10\arcmin$, or 25 pc at the Galactic
Center, which is sufficient to resolve the disk but inadequate for 
individual clouds or the complicated structures in the Galactic
nucleus. The angular resolution of $\la 1 \arcmin$  attained by the  
 synthesis surveys presented elsewhere in this book allows
study of structure in the warp in the first and fourth quadrants,
though any single spectrum at this resolution may be  dominated
by small clouds in self-absorption and not by   large-scale
features.  But the current synthesis surveys do not have the sensitivity
 to detect most halo gas, or most high-velocity clouds, 
or even to detect the disk in some high-latitude directions.  
There will be 
a continued need for different data sets, with different resolution  
and sensitivity, for different investigations.  

A list of major HI emission surveys
is given in Burton (1992) and  Binney \& Merrifield (BM98), 
 to which can be added the southern counterpart to the 
Leiden-Dwingeloo survey (Arnal et al. 2000),
 the 140 Foot Telescope survey of the northern Galactic Plane (Lockman \&
Murphy in preparation),
the HIPASS, HIJASS, CGPS, SGPS, VGPS, and IGPS discussed elsewhere in 
this volume.

\subsection{More Computing, More Ideas}

The data from all major HI surveys can now be stored and manipulated
by relatively inexpensive computers -- the type 
accessible to most astronomers, indeed, to most people
with any kind of education.  Galactic HI studies have, in
principle, become democratized, and those willing to make the effort 
can test their ideas against the data, and possibly even make a
discovery, all in the comfort of their high school classroom, 
 home,  or  pub.  Though all the easy
questions have already been asked, the contrast with 1961 could not be
more pronounced. 

One area where this increase in computing power may 
pay off is in revealing HI features that cross longitude, latitude 
and velocity, and do not appear coherent in the traditional 
projections.  Advanced visualization software may remove their obscurity.

\subsection{Information from Other Species}

The ability of the 21cm  line  to ``see through the dust'', which
made it of such interest to Oort, is now shared by many other 
lines  in the radio (e.g., 
 Dame, Hartmann, \& Thaddeus 2001; Roshi \& Anantharamaiah 2001). 
Fifteen years after Oort's review,  major efforts were underway to 
compare HI with ionized and  molecular gas (Burton 1976).  These days, 
observations  in the optical and UV, though limited to directions of moderate
extinction, contribute increasingly to  Galactic structure studies. 
One recent, striking, result is the detection of the Galactic warp 
in UV absorption lines toward a distant quasar (Savage, Sembach, \& Lu
1995).  The data suggest 
that the  metallicity at the Galaxy's edge falls to 0.1 solar, 
which is the  metallicity of some high-velocity clouds.  
A similar study in another direction shows that the 
abundances in a high-velocity cloud differ from those in  an
intermediate-velocity cloud, implying that the two clouds have very 
different histories (Richter et al.~2001). 
Current  $H\alpha$ telescopes have the resolution in angle and velocity 
of the early HI surveys (Reynolds, Sterling,
\& Haffner 2001) and are  providing information on the relationship
between HI and H$^+$  above the disk (Reynolds et al.~1995). 

These other species may provide critical information to HI studies.
  A measurement of the HI scale height, for example, will be
profoundly misleading if some fraction of the  clouds 
 are ionized at high $|z|$, neutral near the plane, but turn 
 molecular at very low $|z|$,  yet that behavior is quite possible 
in environments where the time scale for a phase transition is 
smaller than the dynamical time (Combes \& Becquaert 1997). 
Another example is high-velocity clouds, which  have long been the 
province of 21cm studies alone.  The discovery of a
cloud which is predominantly ionized (Sembach et al. 1999) shows 
 that even for high-velocity clouds, 
 HI may be giving us just the tip of the iceberg.

\subsection{HI in Absorption}

HI  emission measurements  have made enormous contributions to 
our understanding of the Galaxy, and it  is safe to predict that they 
will still be important in 50 years, for 21cm emission comes from 
varied environments and  can trace subtle phenomena in all directions. 
HI absorption measurements have had a more limited impact.
The 21cm line can be  seen in absorption against 
bright continuum sources (e.g. Radhakrishnan et al. 1972; Garwood \& 
Dickey 1989), and,  sometimes,  in self-absorption against background 
HI emission (e.g. Knapp 1974; Baker \& Burton 1979).  
Self-absorption has attracted recent attention because 
the high-resolution CGPS  has shown that the sky
is peppered with cool clouds (Gibson et al. 2000; see also the 
papers in this volume), but self-absorption requires such special geometry 
that it seems unlikely that it will yield statistically unbiased 
information on Galactic structure.  Absorption against radio 
continuum sources, however,   is another story.

The sensitivity of absorption surveys is discussed in
DL90.  There are hundreds of Galactic HII regions 
which are bright-enough and
compact-enough to yield a good HI absorption spectrum 
in  an hour with existing instruments.  Their
 velocities can be derived from recombination lines, and for some, 
accurate distances can be determined from 
astrometric observations of associated OH or H$_2$O masers.
Pulsars, supernova remnants and some stars can also serve as targets. 
Extragalactic continuum sources can be used  for measurement of the HI
absorption through the entire disk of the Galaxy, analogous to 
emission measurements.

The current synthesis surveys are beginning to give a sense of 
what might be possible in  mapping  the cool HI throughout the Galaxy.  
It is amusing to consider that, with enough data, it may even be profitable to 
resume the old search for the HI spiral arms of the Galaxy, 
only this time not in emission, but in absorption!


\begin{references}

\reference Arnal, E.M., Bajaja, E., Larrarte, J.J., Morras, R., \& 
Poppel, W.G.L. 2000, \aaps, 142, 35
%% A high sensitivity HI survey of the sky at delta <=-25

\reference Baker, P.L. \& Burton, W.B. 1979, \aaps, 35, 129
%% self absorption at Arecibo

\reference Bania, T.M., \& Lockman, F.J. 1984, \apjs, 54, 513
%% Arecibo HI

\reference Binney, J., \& Dehnen, W. 1997, MNRAS, 287, L5
%% The outer rotation curve of the Milky Way

\reference Binney, J., \& Merrifield, M. 1998, Galactic Astronomy, Princeton (BM98)

\reference Blaauw,A., Gum,C.S., Pawsey,J.L., \& Westerhout,G. 1960, \mnras, 121, 123

%%\reference Bland-Hawthorn, J. 1998, in Galactic Halos: a UC Santa Cruz
%%Workshop, ASP Conference Series, Vol 136, p. 113
%% Optical rotation curves beyond the HI cut-off in spirals

\reference Blitz, L. 1993, in Back to the Galaxy, ed. S.S. Holt \&
F. Verter, AIP Conference Proceedings 278, p. 98
%% about the galactic bar

\reference Blitz, L. 1994, in Physics of the Gaseous and Stellar disks of the Galaxy, ed. I. King, ASP
Conference Series 66, p. 1
%% Where is the center of the Milky Way?

\reference Blitz, L., \& Spergel, D.N. 1991a, \apj, 370, 205
%% The shape of the Galaxy

\reference Blitz, L., \& Spergel, D.N. 1991b, \apj, 379, 631
%% direct evidence for a bar at the Galactic center

\reference Briggs, F.H. 1990, \apj, 352, 15
%% rules of behavior for galactic warps

\reference Burton, W.B. 1971, \aap, 10, 76
%% shape of HI profiles and noncircular motions

\reference Burton, W.B. 1976, ARAA, 14, 275
%% Morphology of hydrogen and other tracers in the Galaxy

\reference Burton, W.B. 1988, in Galactic and Extragalactic Radio
Astronomy 2nd edition, ed. G.L. Verschuur, \& K.I.
Kellermann, (Springer-Verlag), p. 295
%% The structure of our Galaxy derived from observations of neutral hydrogen

\reference Burton, W.B. 1992, in The Galactic Interstellar Medium, Saas-Fee Advanced Course 21, ed. D.
Pfenniger, \& P. Bartholdi, (Springer-Verlag), p. 1

\reference Burton, W.B. \& Gordon, M.A. 1978, \aap, 63, 7
%%

\reference Burton, W.B., \& Liszt, H.S. 1978, \apj, 225, 815
%% gas distribution in central region of the galaxy. I. Atomic Hydrogen.

\reference Burton, W.B., \& Liszt, H.S. 1983, \aaps, 52, 63


\reference Burton, W.B., \& Liszt, H.S. 1993, \aap, 274, 765
%% Kinematics of neutral gas in the bulge of the Milky Way


\reference Combes, F., \& Becquaert, J-F 1997, \aap, 326, 554
%% Vertical equilibrium of molecular gas in galaxies

\reference Cox, D.P. \& Reynolds, R.J. 1987, ARAA, 25, 303
%% The local interstellar medium

\reference Dame, T.M. 1993, in Back to the Galaxy, ed. S.S. Holt \&
F. Verter, AIP Conference Proceedings 278, p. 267
%% The distribution of neutral gas in the milky way

\reference Dame, T.M., Hartmann, D., \& Thaddeus, P. 2001, \apj, 547, 792
%% the milky way in molecular clouds: a new complete co survey

\reference Dickey, J.M. 1993, in The Minnesota Lectures on Clusters of Galaxies and Large-Scale
Structure, ed. R.M. Humphries, ASP Conf. Ser. Vol. 39, p. 93
%% Gas as a tracer of Galactic Structure

\reference Dickey, J.M., \& Lockman, F.J. 1990, ARAA, 28, 215 (DL90)
%% HI in the Galaxy

\reference Diplas, A., \& Savage, B.D. 1991, \apj, 377, 126
%%

\reference Englmaier, P., \& Gerhard, O. 1999, \mnras, 304, 512
%% Gas dynamics and large-scale morphology of the milky way Galaxy.

\reference Ewen, H.I., \& Purcell, E.M. 1951, Nature, 168, 356
%% Radiation from Galactic Hydrogen at 1,420 Mc./sec.

\reference Ferrara, A. 1993, \apj, 407, 157
%% Can Galactic HI be radiatively supported?

\reference Florido, E., Battaner, E., Prieto, M., Mediavilla, E., \&
Sanchez-Saavedra, M.L. 1991, MNRAS, 251, 193
%% Corrugations in the discs of spiral galaxies, NGC 4244 and 5023

\reference Freeman, K. C. 2001, in Gas and Galaxy Evolution, ed. J.E. Hibbard, M.P. Rupen, \& J.H. van Gorkom, ASP Conference Series, Vol. 240, p. 301
%% Dark Halos: An HI Perspective

\reference Fux, R. 1999, \aap, 345, 787 
%% 3D self-consistent N-body barred models of the Milky Way

Garwood, R.W., \& Dickey, J.M. 1989, \apj, 338, 841
%% HI absorption survey

\reference Georgelin, Y.M. \& Georgelin, Y.P. 1976, \aap, 49, 57

\reference Gibson, B.K. et al. 2001, \aj, 122, 3280
%% high-velocity cloud complex c:  galactic fuel or galactic waste?

\reference Gibson, S.J., Taylor, A.R., Higgs, L.A., \& Dewdney, P. 2000, \apj, 540, 851
%% cgps HI self-absorption survey

\reference Green, D.A. 1993, \mnras, 262, 327
%% power spectrum analysis of angular scale of Galactic HI ...

\reference Gum, C.S., Kerr, F.J., \& Westerhout, G. 1960, \mnras, 121, 132 

\reference Habing, H.J., 1966, BAIN, 18, 323
%% hvcs and the outer arm

\reference Hartmann, D. \& Burton, W.B. 1997, Atlas of Galactic Neutral Hydrogen (Cambridge Univ.
Press)

\reference Haud, U. 1992, \mnras, 257, 707
%% Dynamics of the Outer Arm high-velocity cloud and the triaxiality of the Milky Way

\reference Hoekstra, H., van Albada, T.S., \& Sancisi, R. 2001, \mnras, 323, 453
%% On the apparent coupling of neutral hydrogen and dark matter in spiral galaxies

%%\reference Howk, J.C. \& Savage, B.D. 1999, \apj, 517, 746
%% Dust in the ionized medium of the Galaxy; 
%% GHRS measurements of AL III and SIII
 
%%\reference Kent, S.M. 1992, \apj, 387, 181
%% rotation curve in bulge from stellar data

\reference Kerr, F.J., Hindman, J.V., \& Carpenter, M.S. 1957, Nature, 180, 677
%% The large-scale structure of the Galaxy

\reference Kerr, F.J. 1962, \mnras, 123, 327 
%%

\reference Kerr, F.J. 1969, ARAA, 7, 39
%% Large scale distribution of hydrogen in the Galaxy

\reference Kerr, F.J. \& Westerhout, G. 1965 in Stars and Stellar Systems,
Vol. 5, ed. A. Blaauw \& M Schmidt, Chicago, p.~167.

\reference Knapp, G.R. 1974, \aj, 79, 527
%% self-absorption

\reference Knapp, G.R., Tremaine, S.D., \& Gunn, J.E. 1978, AJ, 83, 1585
%% HI distribution outside the solar circle 

\reference Kuijken, K., \& Tremaine, S. 1994, \apj, 421, 178
%% On the ellipticity of the Galactic Disk

\reference Kuijken, K., \& Garcia-Ruiz, I. 2001, in Galaxy Disks and Disk Galaxies, ed. J.G. Funes, \&
E.M. Corsini, ASP Conference Series Vol. 230, p. 401
%% Galactic Disk Warps

\reference Kulkarni, S.R., Blitz, L., \& Heiles, C. 1982, \apj, 259, L63
%% Atomic hydrogen in the other milky way

\reference Kuntz, K.D. \& Danly, L. 1996, \apj, 457, 703
%% Intermediate-velocity gas in the north Galactic hemisphere: HI studies

\reference Liszt, H.S. 1992, in The Center, Bulge, and Disk of the Milky Way, ed. L. Blitz, Kluwer, 111
%% HI in the inner Galaxy

\reference Liszt, H.S. \& Burton, W.B. 1978, \apj, 226, 790
%% gas distribution in central region of the galaxy II CO 

\reference Liszt, H.S. \& Burton, W.B. 1980, \apj, 236, 779 
%%% 3 kpc arm as bar

\reference Liszt, H.S. \& Burton, W.B. 1996, in Unsolved Problems of the 
Milky Way, ed. L. Blitz \& P. Teuben, Kluwer, p. 297.
%% Motions and deformations of the inner-galaxy Neutral gas layer

\reference Lockman, F.J. 1977, \aj, 82, 408
%% location of Population-I type objects with respect to the Galactic plane

\reference Lockman, F.J. 1984, \apj, 283, 90
%% HI halo in the inner Galaxy

\reference Lockman, F.J., Hobbs, L.M., \& Shull, J.M. 1986, \apj, 301, 380
%% halo HI

\reference Lockman, F.J. 1988, in The Outer Galaxy, ed. L. Blitz \& F.J. Lockman, (Springer), p. 79
%% Rotation and the Outer Galaxy

\reference Lockman, F.J., \& Gehman, C.S. 1991, \apj, 382, 182
%% 

\reference Lockman, F.J., Murphy, E.M., Petty-Powell, S., \& Urick, V.J. 2002, \apjs, (in press)
%% very sensitive 21cm survey for Galactic high-velocity HI 

\reference Malhotra, S. 1995, \apj, 448, 138
%% vertical distribution and kinematics of hi and mass models of the Galactic disk

\reference Merrifield, M.R. 1992, AJ, 103, 1552
%% Rot curve of Milky Way to 2.5R0 from the thickness of the HI layer

\reference Merrifield, M.R. 1993, \mnras, 261, 233 
%% The kinematics of face-on disc galaxies, and the nature of the Galactic HI layer

\reference Muller, C.A., \& Oort, J.H.  1951, Nature, 168, 357
%% The interstellar hydrogen line at 1420 mc/sec and an estimate of Galactic rotation

\reference Olling, R.P., \& Merrifield, M.R. 2000, MNRAS, 311, 361
%% two measures of the shape of the Milky Way's dark halo

\reference Oort, J. 1961a, in The Distribution and Motion of Interstellar Matter in Galaxies, ed. L. Woltjer,
(Benjamin:New York), p. 3
%%  

\reference Oort, J. 1961b, ibid. p. 71
%%   HI in the Galactic halo

\reference Oort, J.H. 1977, ARAA, 15, 295
%% review of the Galactic center

\reference Oort, J.H., \& Rougoor, G.W., 1960, \mnras, 121, 171
%% location of the galactic center

\reference Peters, W.L. III 1975, \apj, 195, 617.

\reference Pfenniger, D., Combes, F., \& Martinet, L. 1994, \aap, 285,
79 
%% Is dark matter in spiral galaxies cold gas? I. Observational ...

\reference Putman, M.E. \& Gibson, B.K. 1999, PASA, 16, 70
%% First results from the Parkes Multibeam High-velocity cloud survey

\reference Quiroga, R. 1974, \apss, 27, 323
%%

\reference Radhakrishnan,V., Goss,W.M., Murray,J.D., \& Brooks,J.W. 1972, \apjs, 24, 49
%% Parkes HI absorption survey

\reference Reynolds, R.J., Tufte, S.L., Kung, D.T., McCullough, P.R., \& 
Heiles, C. 1995, \apj, 448, 715
%% a comparison of diffuse ionized and HI away from the galactic plane:
%% H-alpha emitting HI clouds

\reference Reynolds, R.J., Sterling, N.C., \& Haffner, L.M. 2001, \apj, 558, L101
%% Detection of a large arc of ionized hydrogen far above ... a superbubble blowout into halo

\reference Richter, P. et al. 2001, \apj, 559, 318
%% Diversity of high and IV clouds: Complex C versus IV Arch

\reference Roshi, D.A., \& Anantharamaiah, K.R. 2001, \apj, 557, 226
%% Hydrogen recombination lines near 327 MHz; physical prop and origin

\reference Sancisi, R. 1996, in The Westerbork Observatory, Continuing 
Adventure in Radio Astronomy, ed.~E.~Raimond \& R.~Genee, Kluwer, p.~143
%% Dark Matter and neutral Hydrogen in Spiral Galaxies

\reference Sancisi, R.,  Fraternali, F., Oosterloo, T., \& van Moorsel,
G. 2001, in Gas and Galaxy Evolution, ed. J.E. Hibbard, M.P. Rupen, 
\& J.H. van Gorkom, ASP Conference Series, Vol. 240, p.~241
%% The HI halos of spiral galaxies

\reference Savage, B.D. 1990, in The Evolution of the Interstellar Medium, 
ed. L. Blitz, ASP Conf. Ser. 12, p.~33
%% Properties of the ISM: gas in the halo

\reference Savage, B.D., Sembach, K.R., \& Lu, L. 1995, \apj, 449, 145
%%site line to H1821+643

\reference Sembach, K. 1997, in Galactic Halos: A UC Santa Cruz Workshop, ed. D. Zaritsky, ASP
Conference Series Vol. 136, p. 97
%% Elemental abundances in Milky Way Halo gas

\reference Sembach, K.R., Savage, B.D., Lu, L., \& Murphy, E.M. 1999, \apj, 515, 108
%% highly-ionized high-velocity clouds

\reference Sembach, K., Howk, J.C., Savage, B.D., \& Shull, J.M. 2001, \aj, 121, 992
%%  FUSE observations of ... in the leading arm of the Magallenic stream

\reference Shane, W.W. 1972, \aap, 16, 118
%% HI in the inner region of the Galaxy (model for the 3-kpc arm)

\reference Simonson, S.C. III 1970, \aap, 9, 163
%% problems in Galactic spiral structure: an account of a spiral workshop

\reference Simonson, S.C. III 1971, Transcription of ``A Spiral Workshop Held at the Univ. of
Maryland, February 16-17 1970",  (Astronomy Program: Univ. Md.)

\reference Spicker, J. \& Feitzinger, J.V. 1986, \aap, 163, 43
%% Are there typical corrugation scales in our Galaxy?

\reference Taylor, J.H., \& Cordes, J.M. 1993, \apj, 411, 674
%% model for ionized gas distribution in Galaxy

\reference Voskes, T., \& Burton, W.B. 1999, in New Perspectives on
the ISM, ASP
Conf. Ser. 168, ed. A.R. Taylor, T.L. Landecker \& G. Joncas, p. 375
%% The large-scale structure of the outer-Galaxy HI layer


\reference Wakker, B.P., \& van Woerden, H. 1997, ARAA, 35, 217
%%  High-velocity clouds

\reference Weaver, H. 1974, in Highlights of Astron., ed.~G. Contopoulos, (Reidel), p.~423
%% Space distribution and motion of the local HI gas

\reference Weaver, H. \& Williams, D.R.W. 1973, \aaps, 8, 1
%% Berkeley low-latitude survey -- description and spectra

\reference Weiner, B.J., \& Sellwood, J.A. 1999, \apj, 524, 112
%% The properties of the galactic bar implied by gas kinematics in milky way 

\reference van Woerden, H., Schwarz, U.J., Reynier, F.P., Wakker,
B.P., \& Kalberla, P.M.W. 1999, Nature, 400, 138
%% Confirmed location in the Galactic halo for the hvc chain A

\reference Wouterloot, J.G.A., Brand,J., Burton,W.B., \& Kwee,K.K. 1990, \aap, 230, 21
%% IRAS sources in the Galactic warp

\end{references}
\end{document}